\documentclass[conference]{IEEEtran}
\ifCLASSINFOpdf
\else
\fi

\usepackage{graphicx}
\usepackage{amsmath, amsthm, amssymb,textcomp}
\usepackage{algorithm} 
\usepackage{cite}
\usepackage{booktabs}
\usepackage{url}
\usepackage{subcaption} 
\usepackage{algorithmicx}
\usepackage{algpseudocode}
\usepackage{paralist}
\usepackage{color}
\usepackage{relsize}
\usepackage{mathptmx} 
\usepackage{stmaryrd}
\usepackage{epsfig} 
\usepackage{mathptmx} 
\usepackage{amsmath,bm} 
\usepackage{amssymb}  
\usepackage{listings}
\usepackage{color}
\usepackage{mathtools}

\usepackage{xspace}
\usepackage{soul,xcolor}
\usepackage{float}

\renewcommand{\r}{}
\renewcommand{\b}{}

\DeclarePairedDelimiter\floor{\lfloor}{\rfloor}

\newcommand{\eco}{\text{$\mathbb{EN}$}}
\newcommand{\encPHE}[1]{\llbracket #1\rrbracket}
\newcommand{\decPHE}{\texttt{DEC}}

\newcommand{\frm}{\textbf{\emph{CryptoImg}}} 

\begin{document}
%
\title{\frm : Privacy Preserving Processing Over Encrypted Images}

\author{\IEEEauthorblockN{M. Tarek Ibn Ziad\IEEEauthorrefmark{1},
Amr Alanwar\IEEEauthorrefmark{2},
Moustafa Alzantot\IEEEauthorrefmark{2}, and
Mani Srivastava\IEEEauthorrefmark{2}}
\IEEEauthorblockA{\IEEEauthorrefmark{1}Ain Shams University, Cairo, Egypt}
\IEEEauthorblockA{\IEEEauthorrefmark{2}University of California, Los Angeles, Los Angeles, California, USA}
Email: mohamed.tarek@eng.asu.edu.eg, \{alanwar, malzantot, mbs\}@ucla.edu}

%


\maketitle



%

\vspace{10pt}
\begin{abstract}

Cloud computing \r{services provide }\b{a scalable solution} for the storage and processing of images and multimedia files. However, concerns about privacy risks prevent users from sharing their personal images with third-party services. 
In this paper, we describe the design and implementation of {\frm}, an open source library\footnote{Source at https://github.com/TarekIbnZiad/CryptoImg} of modular privacy preserving image processing operations over encrypted images. \r{By using homomorphic encryption, {\frm} allows the users to delegate their image processing operations to remote servers without any privacy concerns.} \r{Currently, {\frm} supports} a subset of the most frequently used image processing operations such as image adjustment, spatial filtering, edge sharpening, histogram equalization and others. 
We implemented our library as an extension to the popular computer vision library \texttt{OpenCV}. {\frm} can be used from either mobile or desktop clients. Our experimental results demonstrate that {\frm} is efficient 
while performing operations over encrypted images with negligible error and reasonable time overheads on the supported platforms.

\end{abstract}

\section{Introduction}
\let\thefootnote\relax\footnote{This research is funded in part by the National Science Foundation under awards  CNS-1136174 and CNS-1329755, and by the Center for Excellence for  Mobile Sensor Data-to-Knowledge under National Institutes of Health grant \#1U54EB020404. The U.S. Government is authorized to reproduce and distribute reprints for Governmental purposes notwithstanding any copyright notation thereon. The views and conclusions contained herein are those of the authors and should not be interpreted as necessarily representing the official policies or endorsements, either expressed or implied, of NSF, NIH, or the U.S. Government.}
Cloud computing is one of the fastest growing technologies. Gartner research selected cloud computing among the top 10 strategic technology trends in 2015. Software-as-a-Service (SaaS) is a class of cloud computing that allows thin clients, such as mobile devices or web browsers, to make use of centrally hosted software services on demand. During the past few years, there has been a proliferation of commercial SaaS solutions for various application domains including image editing. For example, services like Adobe Creative Cloud~\cite{adobecloud}, and Pixlr~\cite{pixlir} allow the user to upload pictures from her personal computer or mobile device in order to apply different \r{image enhancements} online.


However, image processing in the cloud \r{presents} a serious threat to the user's privacy. A malicious service provider can look into the user private photos in order to discover sensitive information such as identity, friends, visited places, etc.
As privacy is a crucial issue for end users, mitigating privacy concerns is necessary to increase the adoption of online image processing services. 


In this paper, we present {\frm}, a library of modular image processing operations over encrypted images. 
We implemented our operations by extending the \texttt{OpenCV} library and employing the Paillier cryptosystem~\cite{ref:pail}. 
\b{The major enhancement, which {\frm} introduces as compared to previous work in the field, is that {\frm} can efficiently perform the needed computations with minimal overhead, while guaranteeing the secrecy of private images. {\frm} omits the need for multiple non-collided servers~\cite{journal:Lathey15}.} Also, {\frm} supports different operations including image adjustment, spatial filtering, edge sharpening, edge detection, morphological operations, and histogram equalization over encrypted images. \b{To the best of our knowledge, we are the first to support secure morphological operations besides other image processing operations in one package.} 

Recently, Lathey and Atrey~\cite{journal:Lathey15} introduced a privacy-preserving method for image processing based on Shamir's Secret Sharing (\textbf{SSS}) scheme~\cite{journal:Shamir79}. This method distributes the image enhancement task among multiple servers to ensure privacy. Their solution supports a number of low-level image processing tasks carried out on encrypted images, such as spatial filtering, anti-aliasing, edge enhancement, and dehazing. Although this approach allows performing both addition and multiplication operations over encrypted data, the security of this model is guaranteed only if the computation is distributed over $n$ ($>$1) entities with no more than $k$ among them are colluding. This model is impractical, as it requires non colluding servers and thus provides only weak security guarantees. Moreover, they employed different pre-processing for each secure operation. Therefore, a sequence of secure operations can not be done without decryption and re-encoding. 



The rest of the paper is organized as follows: Section~\ref{sec:pbForm} defines the problem and threat model. Section~\ref{sec:backgrnd} provides a brief background about Paillier cryptosystem and floating-point (FP) encoding technique. It also summarizes the related work. Section~\ref{sec:op} describes our proposed secure operations in details. Section~\ref{sec:expr} provides our experimental evaluation results. Finally, Section~\ref{sec:conc} concludes the paper.
\section{Problem Statement And System Architecture} \label{sec:pbForm}

In this section, first we define the problem statement and threat model. Later, we describe the system architecture of {\frm}.

\subsection{Problem Statement}
We study the problem of protecting the confidentiality of private images against third-party services performing image processing. Our threat model assumes that the clients trust their own hardware and locally-running software programs, but they do not trust third-party remote servers. Although the pressure of market competition forces service providers to perform the requested image enhancement operations correctly, these servers might threaten the user's privacy by abusing the given images to uncover private information for their own business interest. By \r{giving} the server \r{access to nothing more than} encrypted images, our system is secure under the ``honest-but-curious" adversary model. 

\r{We rely on the Paillier cryptosystem which is provably secure using the hardness of \textit{decisional composite residuosity assumption}~\cite{ref:pail}. This means that it is infeasible for any attacker to break the encryption unless he has an efficient algorithm that can solve a family of problems that are computationally intractable. 
Compared to previous work in image processing over encrypted images, our solution } provides better security guarantees than the model adopted by~\cite{conf:Hu06,journal:Lathey15}, which requires more than one server and becomes insecure against colluding servers.

\subsection{System Architecture}\label{sub:setup} 

\begin{figure}[tb]
\centering
\includegraphics*[width = 0.45\textwidth]{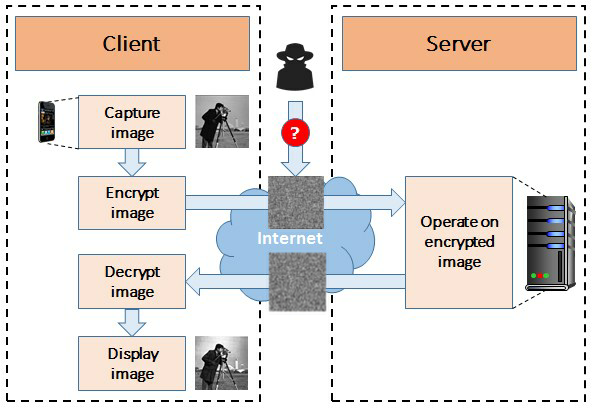}
\caption{System Architecture of {\frm}.}
\label{fig:overallSystem}
\end{figure}

As shown in Fig.~\ref{fig:overallSystem}, {\frm} consists of two parties: client and server. The client represents either an individual personal computer (PC) or mobile device (Mob), while the server is a powerful system offering processing and storage services over the cloud. The client owns private image data and desires to make use of the server image processing services, while keeping the confidentiality of the submitted image against unauthorized access. To achieve this goal, the client encrypts the image before submitting it to the server. Using the homomorphic encryption (HE) properties of Paillier cryptosystem, the server can perform operations over the encrypted image without revealing the source plain-image. The output encrypted image is sent back to the client to decrypt and display the processed image.


\section{Preliminaries and Related work} \label{sec:backgrnd}


This section provides a brief background about Paillier cryptosystem, discusses the used encoding scheme, and summarizes the related work. 

\subsection{Paillier Cryptosystem}
HE is a form of encryption which permits secure computations over encrypted data. 
We denote the encryption of message $m$ using the public key $pk$ as $\encPHE{m}$. Paillier cryptosystem is an additive HE scheme as it provides a public-key operation $\oplus_{z}$ over two encrypted integers which is equivalent to their plain-text addition, as shown in \eqref{eqn:homo}. It also supports a self-blinding operation $\otimes_{z}$ which allows multiplication of encrypted integer by a plaintext scalar $d$, as shown in \eqref{eqn:blnd} $\forall {m_1},{m_2} \in \mathbb Z_{n}\!$. 
\begin{align}
\decPHE(\encPHE{m_1} \oplus_{z} \encPHE{m_2})&= \decPHE( (\encPHE{m_1} \times \encPHE{m_2}) mod\: n^2)  \nonumber\\ &=(m_1 + m_2)\: mod\:n \label{eqn:homo} \\
\decPHE( \encPHE{m_1} \otimes_{z} d) &= \decPHE( \encPHE{m_1}^d \:mod\: n^2 ) \nonumber\\ &= (m \times d )\: mod\: n
\label{eqn:blnd}
\end{align} 



\subsection{From Integers to Floating Point (FP) Numbers} \label{subsec:encode}
Paillier cryptosystem is defined over a group of positive integers $\mathbb {Z}_{n}$, while in practice many operations should happen over real numbers. Therefore, an encoding function \eco{} with minimal quantization error is needed in order to perform secure computation over FP numbers. We define $\phi_{add}{}$ and $\phi_{mul}{}$ as the error introduced due to addition and multiplication operation, as shown in \eqref{eqn:enc_add} and \eqref{eqn:enc_mul}, respectively. Optimal encoding mechanism should have $\phi_{mul}{}  =\phi_{add}{} = 0$. Prior work over encrypted data represents FP numbers through multiplying by a large scaling factor as done in \cite{journal:Shortell15}. However, this representation has $\phi_{mul}{}$ equals the scale factor after each multiplication operation. Thus, it can not be used with arbitrary number of multiplication operations over FP numbers. 
\begin{align}
\phi_{add}{} := abs(\:  \eco( m_1 + m_2 ) - ( \eco(m_1) + \eco(m_2)\: ) ) \label{eqn:enc_add}  \\
\phi_{mul}{} := abs(\:  \eco( m_1 \times m_2 ) - ( \eco( m_1 ) \times \eco( m_2 )\: ) ) \label{eqn:enc_mul}
\end{align} 

Therefore, we have chosen to use the same approach developed by Google's Encrypted BigQuery Client \cite{ref:pailencGoogle}, which represents FP number by a mantissa~$m$ and a non-positive exponent~$e$. A FP number in plaintext is represented by pair ($m$, $e$). In encrypted domain, FP number is represented by a pair of an encrypted mantissa using paillier cryptosystem and an unencrypted exponent ($\encPHE{m}$, $e$). 
Self blinding and additive homomorphic over floats are denoted by $\otimes$ and $\oplus$, respectively. By using the addition and multiplication primitives (~$\oplus_{z}$, $\otimes_{z}$~) of the Paillier cryptosystem, we can perform FP numbers addition and multiplication, as shown in Protocol~\ref{pro:flo}. \b{Also, signed numbers are handled by assigning the ranges $[0,n/3]$ and $[n/3,2n/3]$ for positive and negative numbers, respectively, Whereas the remaining range $(2n/3,n)$ is used for overflow detection. Subtraction accordingly over encrypted floats is denoted by~$\ominus$.}
\begin{algorithm}
\floatname{algorithm}{Protocol}  
\caption{Secure FP Numbers Processing.}
\label{pro:flo}
\begin{algorithmic} [1]
\Statex {\textbf{Multiplication:}} $\encPHE{c} = a \otimes \encPHE{b}$
\Statex $\quad \encPHE{m_{c}} = m_a \otimes_{z} \encPHE{m_b}$
\Statex $\quad e_{c} =  e_a + e_b$
\Statex {\textbf{Addition:}} $\encPHE{c} = \encPHE{a} \oplus \encPHE{b}$
\Statex $\quad $\textbf{ if } $e_a \leq e_b$
\Statex $\quad \quad \encPHE{m_c} = \encPHE{m_a} \oplus_{z} (Base^{e_b - e_a} \otimes_{z} \encPHE{m_b}), \;  e_c =  e_a$
\Statex $\quad $\textbf{ if } $e_a > e_b$
\Statex $\quad \quad \encPHE{m_c} = \encPHE{m_b} \oplus_{z} (Base^{e_a - e_b} \otimes_{z} \encPHE{m_a}), \;  e_c =  e_b$
\end{algorithmic}
\end{algorithm}

\subsection{Related Work} \label{sec:relwrk}


Recently, various privacy preserving algorithms using HE have emerged in different domains including: information retrieval, data mining, and image processing.
Shortell and Shokoufandeh addressed the problem of privacy-preserving image processing by using fully homomorphic encryption (FHE) to process the data while encrypted~\cite{journal:Shortell15}. They used their solution to implement brightness/contrast filter. Also, they extended FHE to support FP numbers via multiplying each value by a factor of $10^d$, where $d$ depends upon the precision of the desired decimal digits up to which we want to process the FP numbers. However, the reported execution time was $15$ minutes on a scaled down image and three hours on the original image. 


\b{Hu \textit{et al.}~\cite{journal:Hu16} proposed a double-cipher method to implement nonlocal means (NLM) denoising over encrypted images. As the NLM operation includes exponentiation, which is a non linear operation, the authors encrypted the plain image with two different cryptosystems before sending to the cloud. The first one was the Paillier scheme, in order to enable the mean filter, and the other was obtained by a distance-preserving transform, in order to enable the nonlocal search. However, their proposed method had higher communication overhead, due to outsourcing two different ciphers for every image. They also enabled only a single type of image processing operations.} 

Moreover, secure multi-party computation (SMC) has been utilized to protect privacy of outsourced images. Hu \textit{et al.} implemented two secure linear filtering protocols~\cite{conf:Hu06}. The first one relied on a combination of rank reduction and random permutation, whereas the second one is based on random perturbation with the help of a third party entity. In the context of secure image retrieval, Zhang \textit{et al.} proposed a secure image retrieval method for cloud computing, which is implemented based on content-based image retrieval (CBIR) framework~\cite{conf:Zhang14}.

%

Hsu \textit{et al.} proposed a privacy-preserving realization of the scale-invariant feature transform (SIFT) method based on Paillier cryptosystem~\cite{journal:Hsu12}. However, their proposed method introduced errors due to the rounding operation in their Gaussian filter coefficients, which were adjusted as integers because their Paillier cryptosystem can only operate in the integer domain. We handle this issue by using appropriate encoding technique.


\section{Secure Operations In Encrypted Domain} \label{sec:op}

The following subsections give details about the supported image processing operations by \frm. 
\subsection{Secure Image Adjustment} 
Image enhancement is done by applying transformation $T$ on an image $I$, which produces the resultant image $R$. We denote the individual pixels values in images~$I$ and~$R$ by~$i$ and~$r$, respectively. Therefore, \r{the relationship between input pixels~$i$ and output pixels~$r$} can be represented by $r = T(i)$. 

{\frm} supports \textit{brightness control} and \textit{image negation}. For \textit{brightness control}, the client requests to adjust the brightness of his image by adding value~$v$, encrypting it along with the image pixels using his public key, $pk$, and sends both the encrypted value $\encPHE{v}$ and encrypted image $\encPHE{I}$ to the server. \r{The server computes the encrypted values of output pixels for all pixels in the image using $\encPHE{r} = \encPHE{i} \oplus \encPHE{v}$. Then, the server} sends the encrypted image back to the client who will decrypt using its secret key $sk$.

Furthermore, {\frm} supports secure \textit{Image negation} \r{where the server computes the encrypted output pixel according to $\encPHE{r} = \encPHE{L - 1} \ominus \encPHE{i}$, for all pixels} in input image with grey levels in the range $[0, L-1]$.



\subsection{Secure Noise Reduction} 
Noise reduction and anti-aliasing operations are essential for many applications like medical, and remote sensing images processing. Smoothing filter in spatial domain is very common operation for anti-aliasing and noise removal, which is equivalent to a low pass filter (LPF) applied in the frequency domain. We denote the output image by $I_{spt}$ whose individual pixels~$(u,v)$ are computed by performing average filter represented in \eqref{eqn:Avg}. The filter $f(u,v)$ is applied first to $m \times n$ patch around~$(u,v)$ pixel, then the intensity values of this patch are averaged.
\begin{align}
\encPHE{I_{spt}(u,v)} = \frac{1}{m \times n} \otimes \sum_{\substack{u = 1, v = 1} }^{m,n} f(u,v) \otimes \encPHE{I(u,v)}
\label{eqn:Avg}
\end{align} 
\b{The challenging part in mapping the average filtering operation to encrypted domain (ED) is how to map the division operation, which may result in a non integer result. As the original Paillier cryptosystem supports only operations over integers, we used our encoding technique, described in sub section~\ref{subsec:encode}. It enables us to multiply by the FP term $1/(m \times n)$.}
Furthermore, arbitrary spatial filter masks can be applied in~\eqref{eqn:Avg}, as we do not restrict the filter value to be positive integers. On the other hand, authors in \cite{journal:Shortell15} did not support negative value in the filter mask.

\subsection{Secure Edge Detection And Sharpening} 

Edge detection is an extremely important step facilitating high-level image analysis \cite{Sonka93}. Edges are pixels where image brightness changes abruptly, therefore gradient operators are commonly used to discover such pixels in the image. {\frm} supports different kind of edge detection operators as Prewitt, Sobel, Robinson, and Kirsh, which are able to detect edges in different directions. Those operators approximates the first derivative. Client sends the encrypted image to the server associated with the required operator ID. Horizontal kernel $h_1$ and vertical kernel $h_2$ are convoluted with the encrypted image to find encrypted horizontal $\encPHE{G_x}$ and vertical $\encPHE{G_y}$ gradient components as shown in  \eqref{eqn:edgde1} and \eqref{eqn:edgde2}. The client decrypts the resultant to find the gradient magnitude $G = \sqrt{ {G_x}^2 + {G_y}^2 }$ and gradient's direction $\Theta = \operatorname{atan2}\left({ G_y , G_x }\right) $.
\begin{align}
\encPHE{G_x(u,v)} = \sum_{\substack{u = 1, v = 1} }^{m,n} h_1(u,v) \otimes \encPHE{I(u,v)} \label{eqn:edgde1}\\
\encPHE{G_y(u,v)} = \sum_{\substack{u = 1, v = 1} }^{m,n} h_2(u,v) \otimes \encPHE{I(u,v)}
\label{eqn:edgde2}
\end{align} 
Additionally, edge sharpening operation in \cite{journal:Shortell15} can be reformulated, as shown in  \eqref{eqn:sharp} in order to decrease the number of operation in the encrypted domain. Subtracting the blurred image $I_{LPF}$ from the original one removes the low pass frequency component and yields the edge representation of the original image $I$. \b{ A positive constant, $k$, is used to control the amount of sharpening. For high-boost filtering, $k$ is greater than one, while it equals one in case of unsharp masking. $\encPHE{I_{LPF}}$ can be obtained using \eqref{eqn:Avg} using the appreciate mask.}

\begin{align}
\encPHE{I_{shrp}(u,v)}= ((k +1) \otimes \encPHE{I(u,v)})  \ominus  (k \otimes \encPHE{I_{LPF}(u,v)})    \label{eqn:sharp}
\end{align}




\subsection{Secure Morphological Operations} 

Morphological operations represent a relatively separate part of image processing. They are widely used in many applications, such as document analysis, character recognition, industrial inspection, and the analysis of microscopic images in fields like geology, biology, and material science. The basic idea in binary morphological operations, studied in this work, is to probe an image $I$ with a pre-defined shape, called the structuring element $B$ with size $m \times n$ . The main two operations in binary morphology are erosion and dilation. Based on these two operations, more complex morphological operations can be computed, such as opening, closing, and shape decomposition. Protocol \ref{pro:Erosion} describes the secure erosion and dilation operations. The erosion threshold value $T$ equals the number of ones in $B$. Conversely, the threshold value $T$ is equal to 1 to perform dilation.


\begin{algorithm}
\floatname{algorithm}{Protocol}  
\caption{Secure Morphological Operations.}
\label{pro:Erosion}
\begin{algorithmic} [1]
\State {Client sends $\encPHE{I}$ to the server associated with the requested structuring element $B$.}

\State {Server performs $\encPHE{L(u,v)} = \sum_{\substack{u = 1, v = 1} }^{m,n} \encPHE{I(u,v)}$.}
\State {Server sends $\encPHE{L}$ to the client.}
\State {Client decrypts $\encPHE{L}$ using his private key. Then, Image thresholding is applied on $L$ using threshold value $T$.}
\end{algorithmic}
\end{algorithm}

\subsection{Secure Histogram Equalization} 
Histogram equalization is a commonly used operation for contrast enhancement. It aims to create an image with equally distributed brightness levels over the whole brightness scale. As shown in Protocol~\ref{pro:Histogram}, this goal is performed by calculating the cumulative image histogram $H_c$ for the input image. Then, a monotonic pixel brightness transformation $T(p)$ is applied such that the desired output histogram is almost uniform over the whole brightness scale. Original image histogram is denoted by $H$ and its size is $G$. Image size is $w \times \ell$. Intensity level is denoted by $p$. 
\begin{algorithm}
\floatname{algorithm}{Protocol}  
\caption{Secure Histogram Equalization.}
\label{pro:Histogram}
\begin{algorithmic} [1]
\State {Client sends encrypted image histogram $H$.}
\State {Server computes the brightness transformation $T(p)$ as following:} 
\Statex {$\encPHE{H_c(0)} = \encPHE{H(0)}$}
\Statex {$\encPHE{H_c(p)} = \encPHE{H_c(p-1)} \oplus \encPHE{H(p)}, where\:p = 1,2,\cdots G-1$}
\Statex {$\encPHE{T(p)} = (G-1)/(w \times \ell) \otimes \encPHE{H_c(p)}$.} 
\State {Server sends $\encPHE{T}$.}
\State {Client decrypts and applies $T(p)$ on each image pixel.}
\end{algorithmic}
\end{algorithm}




\section{Evaluation} \label{sec:expr}

{\frm} is implemented in C++ using GMP and NTL as an extension to the popular computer vision library \texttt{OpenCV}. We also developed an Android client application, which is implemented in Java. Our implementation of Paillier cryptosystem extends the work of~\cite{Link:Paillier} to introduce the FP support described earlier in SubSection \ref{subsec:encode}. For our experiments, we used a Intel Xeon(R) desktop machine with 8 cores at 2.20~GHz running Ubuntu 64-bit operating system. Our Android client application is installed on Nexus 5 (NX) mobile device, with Quad-core 2.30 GHz Krait 400 CPU. 

The rest of this section provides our {\frm} evaluation in terms of the introduced error, computation time on both client and server, and communication overhead. 
 
\subsection{Visual Output Evaluation} 

We performed a number of experiments to evaluate the performance of different operations supported by \frm. We applied the operations to a number of gray level images from the public CVG-UGR gray level image database~\cite{Link:CVG}; The dimensions of every image is $512 \times 512$ pixels and every pixel is represented by 8 bits. \b{In case of morphological operations, selected binary images from another database~\cite{Link:MPEG7}}.

\begin{figure*}[!t]
\centering
\begin{tabular} {c|c|c|c|c|c|c}
\textbf{Negation} & \textbf{Brightness} & \textbf{LPF} & \textbf{Edges} & \textbf{Sharpening} & \textbf{Dilation} & \textbf{Equalization} 
\\[10pt]
\includegraphics[width = 0.11\textwidth]{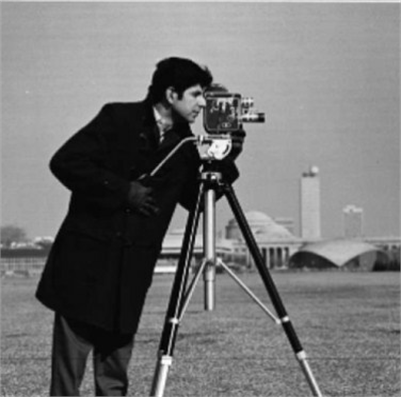} &
\includegraphics[width = 0.11\textwidth]{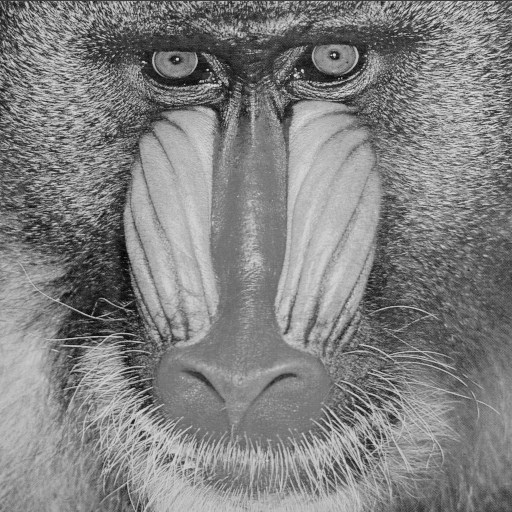} &
\includegraphics[width = 0.11\textwidth]{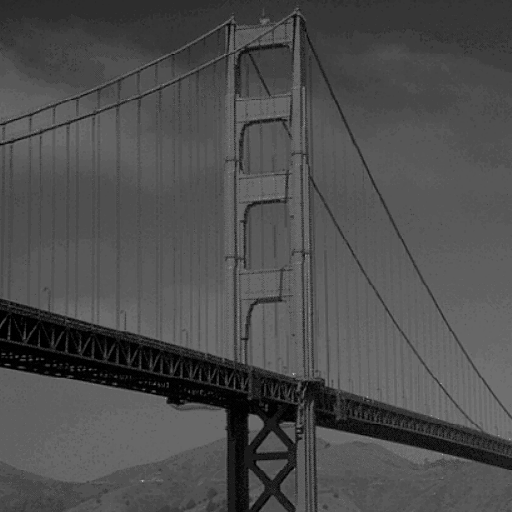} &
\includegraphics[width = 0.11\textwidth]{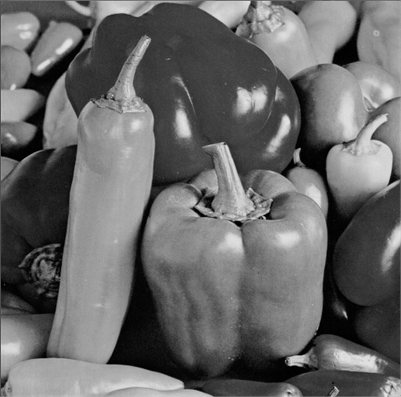} &
\includegraphics[width = 0.11\textwidth]{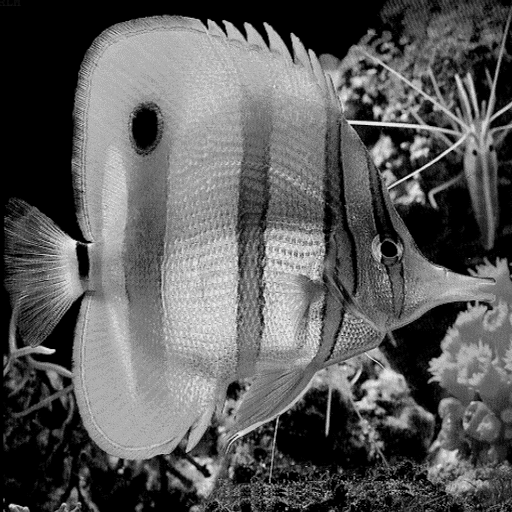} &
\includegraphics[width = 0.11\textwidth]{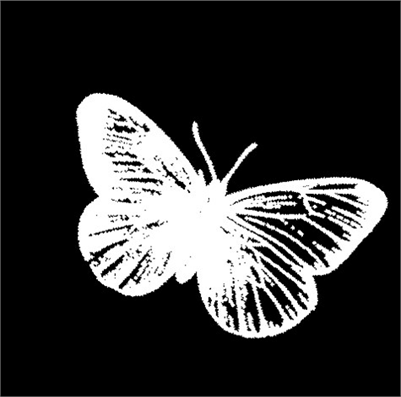} &
\includegraphics[width = 0.11\textwidth]{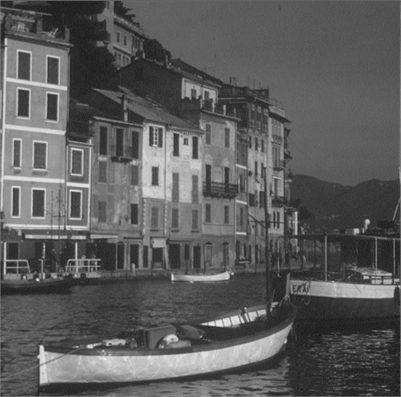} \\
(a) Input & (d) Input & (g) Input & (j) Input & (m) Input & (p) Input & (s) Input \\[8pt]

\includegraphics[width = 0.11\textwidth]{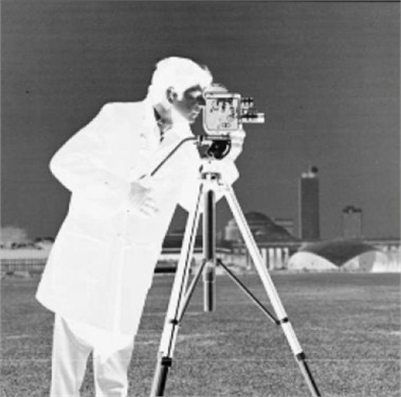} &
\includegraphics[width = 0.11\textwidth]{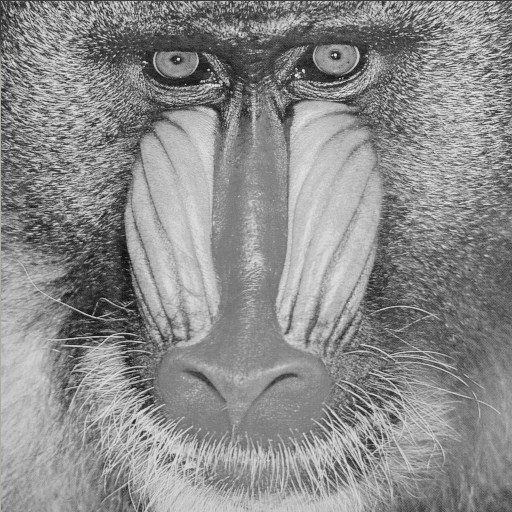} &
\includegraphics[width = 0.11\textwidth]{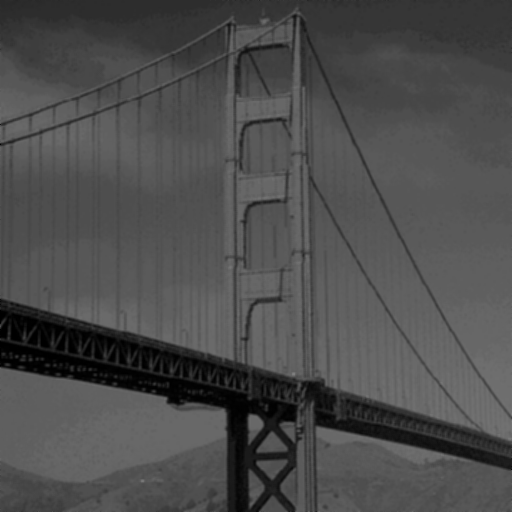} &
\includegraphics[width = 0.11\textwidth]{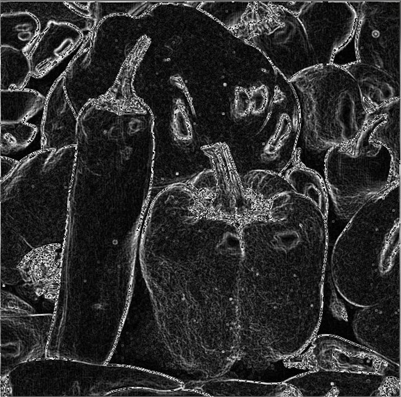} &
\includegraphics[width = 0.11\textwidth]{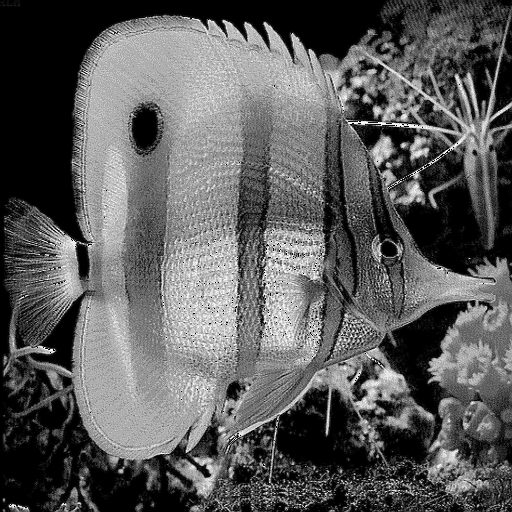} &
\includegraphics[width = 0.11\textwidth]{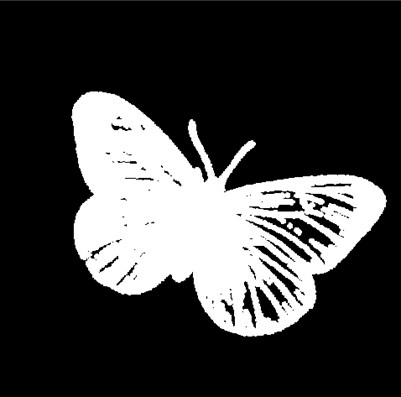} &
\includegraphics[width = 0.11\textwidth]{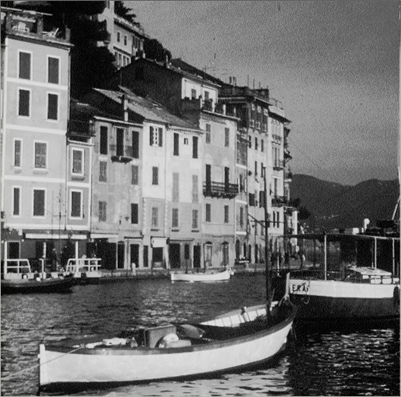} \\
(b) PD Output & (e) PD Output & (h) PD Output & (k) PD Output & (n) PD Output & (q) PD Output & (t) PD Output \\[8pt]

\includegraphics[width = 0.11\textwidth]{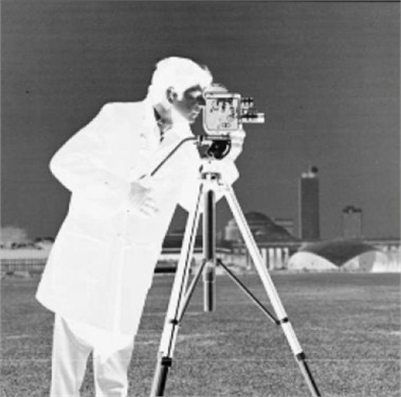} &
\includegraphics[width = 0.11\textwidth]{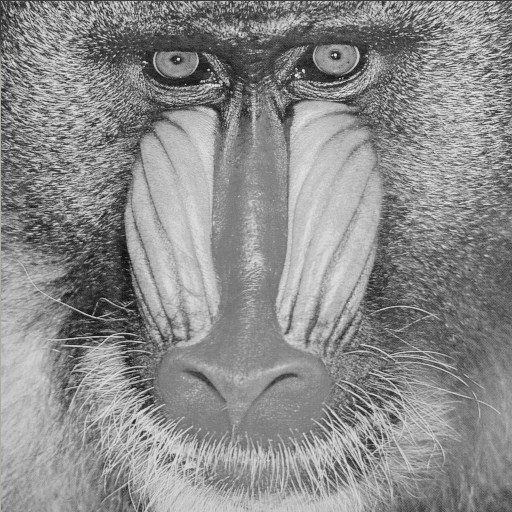} &
\includegraphics[width = 0.11\textwidth]{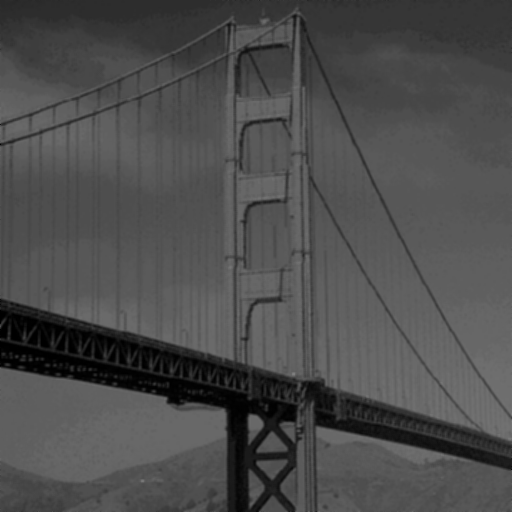} &
\includegraphics[width = 0.11\textwidth]{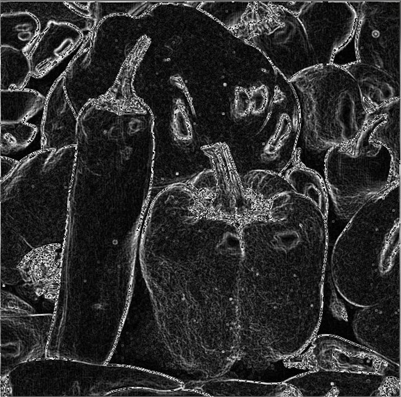} &
\includegraphics[width = 0.11\textwidth]{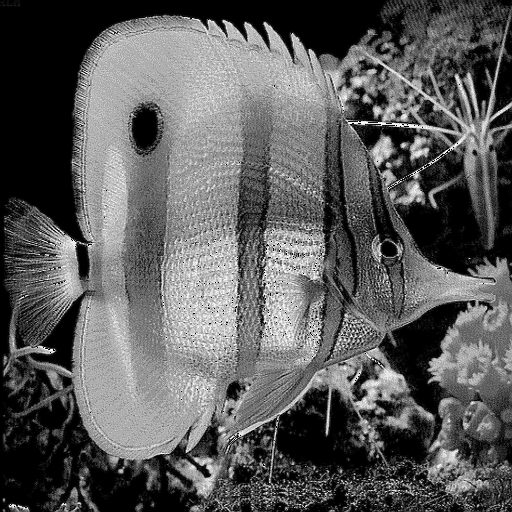} &
\includegraphics[width = 0.11\textwidth]{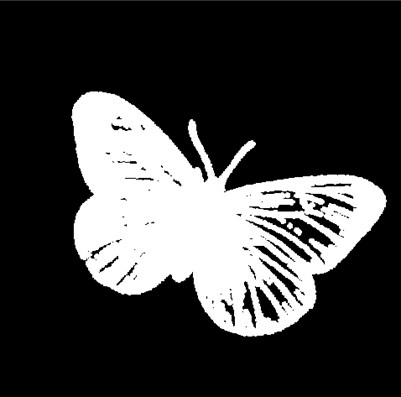} &
\includegraphics[width = 0.11\textwidth]{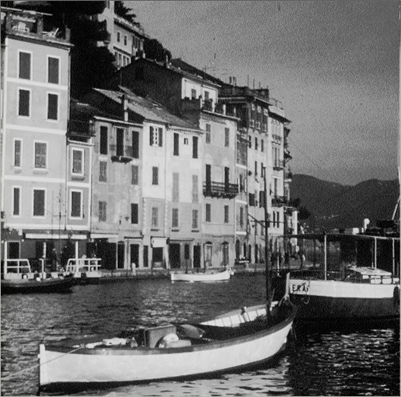}\\
(c) ED Output & (f) ED Output & (i) ED Output & (l) ED Output & (o) ED Output & (r) ED Output & (u) ED Output\\
\end{tabular}
\caption{Visual output evaluation for operations applied in PD and ED using $10^{-8}$ precision level.}
\label{fig:visual_output}
\end{figure*}

Fig.~\ref{fig:visual_output} shows the result of the proposed methods using a precision of $10^{-8}$. The precision determines the exponent of the encoded FP number using $\floor*{\log_{Base} precision}$. Due to space limit, we only show one output for each method. Fig.~\ref{fig:visual_output}-a represents the original images, which is encrypted using user Paillier public key and submitted to the server to obtain image negation. Fig.~\ref{fig:visual_output}-c shows the output after applying image negation in encrypted domain (ED). On the other hand, Fig.~\ref{fig:visual_output}-b shows the output of image negation in the plaintext domain (PD) using normal \texttt{OpenCV} APIs. Fig.~\ref{fig:visual_output}-d through Fig.~\ref{fig:visual_output}-f show the same for brightness adjust. Additionally, Fig.~\ref{fig:visual_output}-i shows the decrypted output after applying secure averaging operation on Fig.~\ref{fig:visual_output}-g using a $3 \times 3$ filter. The visual effect of secure blurring and noise removal is compared with Fig.~\ref{fig:visual_output}-h which is the normal average filter in the PD.

For the sake of testing edge detection techniques, a simple Sobel filter is used to detect edges in Fig.~\ref{fig:visual_output}-j the outputs of the ED and PD are shown in Fig.~\ref{fig:visual_output}-l and Fig.~\ref{fig:visual_output}-k, respectively. On the other hand, edge sharpening with $k=1.0$ is applied on Fig.~\ref{fig:visual_output}-m. Edge sharpening in ED and PD are shown in Fig.~\ref{fig:visual_output}-o and Fig.~\ref{fig:visual_output}-n, respectively. Also, an example for the morphological operations is represented by applying a dilation operation on Fig.~\ref{fig:visual_output}-p. The output in PD and ED are shown in Fig.~\ref{fig:visual_output}-q and Fig.~\ref{fig:visual_output}-r, respectively. Finally, Protocol~\ref{pro:Histogram} is applied on Fig.~\ref{fig:visual_output}-s to perform histogram equalization in ED. The result is shown in Fig.~\ref{fig:visual_output}-u which is compared with PD outputs in Fig.~\ref{fig:visual_output}-t.

By comparing the output of operations in both encrypted and plain domains, we find that all our secure methods introduce zero error except LPF and edge sharpening, which introduce a low error at higher precision. Table \ref{tab:per_err} shows the effect of choosing the precision level in the secure LPF and edge sharpening operations. The error is calculated by comparing the output in PD and ED. Based on that, we choose $10^{-8}$ as a reasonable precision.


\begin{table}[b]
\caption{Precision effect on the introduced error.}
\centering
\normalsize
\resizebox{0.45\textwidth}{!}{
\begin{tabular}{l|rrrr}
    \toprule
    Precision & \multicolumn{2}{c}{Average Error} & \multicolumn{2}{c}{Standard Deviation} \\
\hline\midrule
          & \multicolumn{1}{c}{LPF} & \multicolumn{1}{c}{Sharpening} & \multicolumn{1}{c}{LPF} & \multicolumn{1}{c}{Sharpening} \\
    $10^{-2}$ & 0.768 & 0.644 & 0.471 & 0.485 \\
    $10^{-8}$ & 0.145 & 0.012 & 0.352 & 0.112 \\
    $10^{-10}$ & 0.145 & 0.012 & 0.352 & 0.116 \\
    \bottomrule
    \end{tabular}}%
\label{tab:per_err}
\end{table} 

 
\subsection{Computation Time}

 We used two different implementation for Paillier cryptosystem for PC and Mob. Table~\ref{tab:PaillierTime} shows the computation time that {\frm} takes to encrypt/decrypt images using different key sizes. The encryption/decryption process is done pixel by pixel. Therefore, if the original image size is $n \times n \times 8$ bits and a $k$ bit key is used, the size of the encrypted image would equal approximately $2k \times n \times n$ bits. That represents approximately a $k/4$ expansion factor. Histogram equalization operation does not require the encryption of all pixels. Only the histogram is encrypted, as explained in Protocol~\ref{pro:Histogram}. 


\begin{table*}[tb]
  \centering
  \caption{Execution Time (sec) of the Paillier encryption/decryption of image using different key sizes on both personal computer (PC) and mobile device (Mob) clients. We used $512 \times 512$ image for PC and $256 \times 256$ image for Mob.}
  \resizebox{0.48\textwidth}{!}{
    \begin{tabular}{l|rrrrrr}
    \toprule
    Key Size & \multicolumn{1}{c}{256} & \multicolumn{1}{c}{512} & \multicolumn{1}{c}{1024} & \multicolumn{1}{c}{2048} \\
\hline\midrule
    Encrypt-PC & 23.9164 & 156.905 & 1154.29 & 7670.49 \\
    Decrypt-PC & 1.39223 & 1.93554 & 4.06813 & 9.62313 \\
    \hline
    
    
   Encrypt-Mob & 13 & 73 & 575 & 3701  \\
   Decrypt-Mob & 10 & 48 & 325 & 2268  \\
    \bottomrule
	\end{tabular}}%
	\label{tab:PaillierTime}%
\end{table*}%

    
    

Table~\ref{tab:MethodsTime} shows timing results of running our protocols using a PC or Mob clients with the configuration given in Section~\ref{sub:setup}. For obtaining a high level of security, we set the Paillier key length to of 1024-bits and 2048-bits in all scenarios. Edge sharpening is the most expensive operation, as it needs successive computations. \b{The relatively high cost of the encryption process could be amortized by storing an encrypted version of the image on a cloud storage. The image is encrypted once and could be used as an input for many secure image processing operations.}


\begin{table*}[bt]
  \centering
  \caption{Execution Time (sec) of the proposed operations using 1024-bit and 2048-bit keys on both personal computer (PC) and mobile device (Mob) clients. The server is modeled as the personal computer. We used $512 \times 512$ image.}
  \resizebox{0.72\textwidth}{!}{
    \begin{tabular}{l|rrrrrrr}
    \toprule
    \multicolumn{1}{c}{Operation} & \multicolumn{1}{c}{PD} & \multicolumn{6}{c}{ED} \\
    \multicolumn{1}{c}{} & \multicolumn{1}{c}{} & \multicolumn{2}{c}{Pre-processing} & \multicolumn{2}{c}{Server} & \multicolumn{2}{c}{Post-processing} \\ 
     \multicolumn{1}{c}{} & \multicolumn{1}{c}{} & \multicolumn{1}{c}{PC} & \multicolumn{1}{c}{Mob}& \multicolumn{1}{c}{1024-bit} & \multicolumn{1}{c}{2048-bit} & \multicolumn{1}{c}{PC} & \multicolumn{1}{c}{Mob}\\ 
\hline\midrule
    Negation          & 0.00122  & 0        & 0    & 42.4737  & 137.925  & 0      & 0\\
    Brightness 		  & 0.00108  & 0        & 0    & 0.81994   & 2.39777   & 0      & 0 \\
    LPF               & 0.00763  & 0        & 0    & 180.508  & 609.199  & 0      & 0 \\
    Sobel filter  	  & 0.00642  & 0        & 0    & 147.567  & 482.195  & 0.0012 & 0.0940\\
    Sharpening  	  & 0.00977  & 0        & 0    & 238.257  & 807.528  & 0      & 0\\
    Erosion           & 0.00009  & 0        & 0    & 4.04937  & 10.8085  & 0.0006 & 0.0198 \\
    Dilation 		  & 0.00008  & 0        & 0    & 4.04937  & 10.8085  & 0.0005 & 0.0198\\
    Equalization      & 0.00174  & 0.00182  & 0.177& 0.01446  & 0.04835  & 0.0007 & 0.0290\\
    \bottomrule
    \end{tabular}}%
  \label{tab:MethodsTime}%
\end{table*}

\b{\subsection{Discussion}}

Based on our results, we conclude that performing operations over encrypted images adds more cost in terms of computation time, communication, and storage. The added cost increases as the length of encryption key increases. However, our protocols add minimal computation overhead, which is orders of magnitude less than prior work (e.g.~\cite{journal:Shortell15}) and also minimal communication overhead, only one round between the client and server.

\b{Also, it is worth mentioning that not all image processing operations can be directly implemented in {\frm}. For instance, operations that require sorting/comparison, such as median filtering~\cite{book:Gonzalez}, would require more communication rounds between the client and server. Some gray-scale transformations, such as contrast manipulation, would not be feasible in ED as they rely on the knowledge of the original intensity value of the pixel to be able to map this value to another intensity value. More complicated algorithms, such as SIFT, could be supported in ED at the cost of adding more communication rounds between the client and server, a similar approach was used by~\cite{journal:Hsu12}.}

\section{Conclusion} \label{sec:conc}
In this paper, we introduced {\frm}, a library of modular privacy preserving image processing operations over encrypted images based on the homomorphic properties of Paillier cryptosystem. Secure operations, such as image adjustment, spatial filtering, edge sharpening, edge detection, morphological operations, and histogram equalization, are safely outsourced to third-party servers with no privacy issues. 

We presented how this operations can be implemented with much less time overhead, and single communication round. Moreover, {\frm} can be used from either mobile or desktop clients with low client-side overheads. 
Experiments show the  efficiency of our proposed library. 
\b{In the future, it will be interesting to explore the feasibility of using the current secure operations as building blocks to support more complex algorithms.}



\bibliographystyle{IEEEtran}
\bibliography{ref}
%

\end{document}